\documentclass[sigplan]{acmart}
\setcopyright{none}

\usepackage{epigraph}
\usepackage{booktabs}   
\usepackage{subcaption} 
\usepackage{xspace}
\usepackage{mymacros}
\usepackage{algorithmic}
\usepackage{graphicx}
\usepackage{textcomp}
\usepackage{xcolor}
\usepackage{url}
\usepackage[T1]{fontenc}
\usepackage{array}
\usepackage{xparse}

\usepackage{lang}
\usepackage{mathpartir}
\usepackage{wasysym}
\usepackage{tabularx}
\usepackage{enumitem}

\theoremstyle{plain}
\newtheorem{principle}{Principle}
\newtheorem{corollary}{Corollary}[principle]

\makeatletter
\DeclareRobustCommand{\var}[1]{%
  \ifmmode 
  {\begin{small}{{#1}}\end{small}}
  \else 
  \begin{math}
  {\begin{small}{{#1}}\end{small}}
  \end{math}
  \fi
}
\makeatother

\newcommand{\souffle}{Souffl{\'e}\xspace}

\usepackage{pifont}

\begin{document}


\title[Next-Paradigm Programming Languages]{Next-Paradigm Programming Languages: What Will They Look Like and What Changes Will They Bring?}



\author{Yannis Smaragdakis}
\affiliation{
  \institution{University of Athens}            
}
\email{smaragd@di.uoa.gr}          

\begin{abstract}
  The dream of programming language design is to bring about
  orders-of-magnitude productivity improvements in software development
  tasks. Designers can endlessly debate on how this dream can be
  realized and on how close we are to its realization. Instead, I
  would like to focus on a question with an answer that can be,
  surprisingly, clearer: what will be the common principles behind
  next-paradigm, high-productivity programming languages, and how will
  they change everyday program development? Based on my decade-plus
  experience of heavy-duty development in declarative languages, I
  speculate that certain tenets of high-productivity languages are
  inevitable. These include, for instance, enormous variations in
  performance (including automatic transformations that change the
  asymptotic complexity of algorithms); a radical change in a
  programmer's workflow, elevating testing from a near-menial task to
  an act of deep understanding; a change in the need for formal
  proofs; and more.
\end{abstract}



\maketitle

\vspace{0.5cm}

\epigraph{[W]e've passed the point of diminishing returns. No future language will give us the factor of 10 advantage that assembler gave us over binary. No future language will give us 50\%, or 20\%, or even 10\% reduction in workload.}{\textit{Robert C. Martin~\cite{martin16}}}


\section{Introduction}\label{sec:introduction}

Since the 1950s, high-level programming languages have resulted in
orders-of-magnitude productivity improvements compared to
machine-level coding. This feat has been a great enabler of the
computing revolution, during a time when computer memories and
conceptual program complexity have steadily grown at exponential
rates. The history of computing is testament to language designers'
and implementers' accomplishments: of the 53 Turing awards to present
(from 1966 to 2018) a full 16 have been awarded for contributions to
programming languages or compilers.\footnote{The count is based on
  occurrences of ``programming language(s)'' or ``compiler(s)'' in the
  brief citation text of the award, also including Richard Hamming who
  is cited for ``automatic coding systems'' (i.e., the L2 precursor of
  Fortran).  Notably, the number does not include John McCarthy or
  Dana Scott, who are well-known for languages contributions yet the
  terms do not appear in the citation.}


At this time, however, the next steps in programming language
evolution are hard to discern. Large productivity improvements will
require a Kuhnian \emph{paradigm shift} in languages. A change of paradigm, in the
Kuhn sense, is a drastic reset of our understanding and nomenclature.
It is no surprise that we are largely ineffective at predicting its
onset, its nature, or its principles.

Despite this conceptual difficulty, the present paper is an attempt to
peer behind the veil of next-paradigm programming languages. I happen
to believe (on even days of the month) that a change of paradigm is
imminent and all its technical components are already here. But you do
not have to agree with me on either point---after all, the month also
has its odd days.

Reasonable people may also differ on the possible catalysts of such a
paradigm shift. Will it be machine learning and statistical
techniques~\cite{43146}, trained over vast data sets of code instances?  Will it be
program synthesis techniques~\cite{DBLP:journals/ftpl/GulwaniPS17},
employing symbolic reasoning and complex
constraint solving? Will it be mere higher-level language design
combined with technology trends, such as vast computing power and
enormous memories?

Regardless of one's views, I hope to convince the reader that
there is reasonable clarity on \emph{some} features that
next-paradigm programming languages will have, \emph{if} they
ever dominate. Similarly, there is reasonable clarity on
what changes next-paradigm programming languages will
induce in the tasks of everyday software development.

For a sampling of the principles I will postulate and their
corollaries, consider the following conjectures:

\begin{itemize}[leftmargin=12pt,itemsep=2pt]
  
\item Next-paradigm programming languages will not display on the
  surface the computational complexity of their calculations.  Large
  changes in asymptotic complexity (e.g., from 
  $O(n^4)$ to $O(n^2)$) will be effected by the language
  implementation. The language will not have loops or explicit
  iteration. The programmer will often opt for worse asymptotic
  complexity factors, favoring solution simplicity and countering
  performance problems by limiting the inputs of algorithms (e.g.,
  applying an expensive computation only locally)
  or accepting approximate results.

\item Next-paradigm programming languages will need a \emph{firm
  mental grounding}, in order to keep program development
  manageable. This grounding can include: a well-understood cost model;
  a simple and clear model on how new code can or cannot affect the
  results of earlier code; a natural adoption of parallelism without
  need for concurrency reasoning; and more.
  
\item Development with next-paradigm programming languages will be
  significantly different from current software development.  Minute
  code changes will have tremendous impact on the output and its
  computation cost. Incremental development will be easier.  Testing
  and debugging will be as conceptually involved as coding. Formal
  reasoning will be easier, but less necessary.
\end{itemize}

In addition to postulating such principles, the goal of the paper
is to illustrate them.
I will use examples from real, deployed code, written (often by
me) in a declarative language---Datalog. My experience in declarative
programming is a key inspiration for most of the observations of the
paper. It is also what makes the conjectures of the paper
``real''. All of the elements I describe, even the most surprising,
are instances I have encountered in programming practice. I begin
with this personal background before venturing to further
speculation.

\vspace{0.5cm}
\epigraph{I've seen things, that's why I'm seeing things.}{\textit{---me}}

\section{Where I Come From}

In the past decade, I have had the opportunity to write declarative,
logic-based code of uncommon volume and variety, under stringent
performance requirements. This experience underlies my speculation on
the properties of next-paradigm programming languages.

\paragraph{Declarative code---lots of it.} Most of my research (and
the vast majority of my personal coding effort) in the past decade has
been on declarative program analysis~\cite{10.1007/978-3-642-24206-9_14}. My
group and a growing number of external collaborators have implemented
large, full-featured static analysis frameworks for Java
bytecode~\cite{Bravenboer09}, LLVM bitcode~\cite{10.1007/978-3-662-53413-7_5},
Python~\cite{tensor19}, and Ethereum VM
bytecode~\cite{gigahorse,Grech2018oopsla}. The frameworks have been the
ecosystem for a large number of new static analysis algorithms,
leading to much new research in the area.

These analysis frameworks are written in the Datalog language.
Datalog is a bottom-up variant of Prolog, with similar syntax.
``Bottom-up'' means that no search is performed to find solutions to
logical implications---instead, all valid solutions are computed, in parallel.
This makes the language much more declarative than Prolog: reordering
rules, or clauses in the body of a rule, does not affect the output.
Accordingly, computing all possible answers simultaneously means that
the language has to be limited to avoid possibly infinite
computations. Construction of new objects (as opposed to new
combinations of values) is, therefore, outside the core language and,
in practice, needs to be strictly controlled by the programmer. These
features will come into play in later observations and conjectures.

The Datalog language had been employed in static program analysis long
before our
work~\cite{repsdb,DBLP:conf/aplas/WhaleyACL05,pods/LamWLMACU05,codequest}.
However, our frameworks are
distinguished by being almost entirely written in Datalog: not just
quick prototypes or ``convenient'' core computations are expressed as
declarative rules, but the complete, final, and well-optimized version
of the deployed code, as well as much of the scaffolding of the
analysis.  As a result, our analysis frameworks are possibly the
largest Datalog programs ever written, and among the largest pieces of
declarative code overall. For instance, the Doop codebase \cite{Bravenboer09}
comprises several thousands of Datalog rules, or tens of thousands of
lines of code (or rather, of logical specifications). This may seem like
a small amount of code, but, for logical rules in complex mutual
recursion, it represents a daunting amount of complexity. This
complexity captures core static analysis algorithms, language
semantics modeling (including virtually the entire complexity of
Java), logic for common frameworks and dynamic behaviors, and more.

\paragraph{Emphasis on performance, including parallelism.}
The Datalog code I have contributed to in the past decade aims for
performance at least equal to a manually-optimized imperative
implementation. That is, every single rule is written with a clear
cost model in mind. The author of a declarative rule knows, at least
in high-level terms, how the rule will be evaluated: in how many
nested loops and in what order, with what indexing structures, with
which incrementality policy for faster convergence when recursion is
employed. Optimization directives are applied to achieve maximum
performance. Shared-memory parallelism is implicit, but the programmer
is well aware of which parts of the evaluation parallelize well and
which are inherently sequential. In short, although the code is very
high-level, its structure is anything but random, and its performance
is not left to chance. Maximum effort is expended to encode
highest-performing solutions purely declaratively.

\paragraph{Many domains.} My experience in declarative programming
extends to several domains, although static analysis of programs has
been the primary one. Notably, I have served as consultant, advisor,
and academic liaison for LogicBlox Inc., which developed the Datalog
engine~\cite{Aref:2015:DIL:2723372.2742796} used in Doop until
2017. The company built a Datalog platform comprising a language
processor, JIT-like back-end optimizer, and specialized database
serving as an execution environment. All applications on the platform
were developed declaratively---even UI frameworks were built out of a
few externally-implemented primitives and otherwise entirely
logic-based rules. The company enjoyed a lucrative acquisition,
mostly based on the value of its deployed applications and
customers. All applications were in the domain of retail 
prediction---about as distant from static analysis as one can imagine.
For a different example, an algorithm for dynamic race detection
developed in the course of my
research~\cite{cprace} was implemented
in Datalog and all experiments were over the Datalog
implementation. An imperative implementation would be substantially
more involved and was never successfully completed in the course of
the research.

\paragraph{``Declarative languages aren't good for this.''} A repeat
pattern in my involvement with declarative languages has been to
encode aspects of functionality that were previously thought to
require a conventional (imperative or functional) implementation.  The
possibility of the existence of the Doop framework itself was under
question a little over a decade ago---e.g.,
Lhot\'{a}k~\cite{Lhotak:2006:PAU} writes: \emph{``[E]ncoding all the
  details of a complicated program analysis problem [...] purely in
  terms of subset constraints [i.e., Datalog] may be difficult or
  impossible.''}  In fact, writing all analysis logic in Datalog has
been a guiding principle of Doop---extending even to parts of the
functionality that might be \emph{harder} to write declaratively. Very
few things have turned out to be truly harder---it is quite surprising
how unfamiliarity with idioms and techniques can lead to a fundamental
rejection of an approach as ``not suitable''. Most recently, we got
great generality and scalability benefits from encoding a decompiler
(from very-low-level code) declaratively~\cite{gigahorse}, replacing a previous imperative
implementation~\cite{vandal}---a task that the (highly expert) authors
of the earlier decompiler considered near-impossible.

\paragraph{Design declarative languages.} Finally, I have had
the opportunity to see declarative languages not just from the
perspective of a power user and design advisor, but also from that of
a core designer and implementer~\cite{pql, deal}. This dual view has
been essential in forming my understanding of the principles and
effects of next-paradigm languages.

\vspace{0.5cm}

\newpage

\epigraph{[A]ll programming languages seem very similar to each other. They all have variables, and arrays, a few loop constructs, functions, and some arithmetic constructs.}{\textit{Patrick S. Li~\cite{li16}}}

    
\section{Principles of Next-Paradigm Languages}

Before going on, I will emphasize again that the reader does not need
to agree with me on the usefulness of declarative
languages. Next-paradigm programming languages could be based on any
of several potential technologies---e.g., perhaps on machine learning
and statistical techniques, or on SMT solvers and symbolic reasoning.
Regardless of the technology, however, I think that some elements are
near-inevitable and these are the ones I am trying to postulate as
``principles''. I will illustrate these principles with examples from
declarative programming, because that's the glimpse of the future I've
happened to catch. But other glimpses may be equally (or more) valid.


\subsection{Programming Model: Cost}

\begin{principle}[Productivity and Performance Tied Together]
If a language can give orders-of-magnitude improvements in productivity \\
\textbf{then}
its implementation has the potential for orders-of-magnitude changes in performance.
\label{principle:performance}
\end{principle}

Large variations in both productivity and performance are functions of
a language being \emph{abstract}. Neither is possible with the
current, ultra-\emph{concrete} mainstream languages. If one needs to
explicitly specify ``loops and arrays'', neither large productivity
gains, nor large performance variations are possible. Instead, the
language implementation (or ``compiler'' for short\footnote{A language
  implementation consists of an interpreter or compiler (ahead-of-time
  or just-in-time) and a runtime system. The term ``compiler'',
  although not always accurate, seems to encompass most of the
  concepts we are used to in terms of advanced language
  implementations.}) of a high-productivity, next-paradigm language
will likely be able to effect orders-of-magnitude performance
differences via dynamic or static optimization. For performance
variability of such magnitude, the asymptotic complexity of the
computation will also likely change.

\begin{corollary}
  Programs $\neq$ 
  Algorithms $+$
  Data Structures. \\
  \emph{Instead:}\\
  Compiler(Program) $=$ Algorithms $+$ Data Structures.
\end{corollary}

Programs in next-paradigm languages will likely \emph{not} be the sum
of algorithms and data structures, contradicting Wirth's famous
equality. Instead, programs will be specifications---carefully written
to take into account an execution model that includes a search process
(done by the compiler) over the space of implementations. Major
algorithmic elements and key data structure decisions will be
determined automatically by this search. The compiler will be a mere
function from programs to concrete implementations, consisting of
algorithms and data structures.

\paragraph{Example: Choice of Algorithm.} Language optimization that can affect the
asymptotic complexity of the computation is hardly new. Relational
query optimization is a prime realistic example.\footnote{Relational
  database languages, such as SQL, are a limited form of declarative
  programming. Due to the simplified setting and commercial success,
  many ideas we discuss have originated in that domain.}  In our
setting, we can revisit it, with Datalog syntax, before building
further on it.  A common static analysis rule, responsible for
interpreting calls as assignments from actual to formal arguments, is
shown below:

\noindent\begin{minipage}{\columnwidth}
\begin{datalogcode}
Assign(formal, actual) :-
  CallGraphEdge(invocation, method),
  FormalParam(index, method, formal),
  ActualParam(index, invocation, actual).
\end{datalogcode}
\end{minipage}

The logic just says that if we have computed a call-graph edge from
instruction \sv{invocation} to a \sv{method}, then the $i$-th
(\sv{index}) actual argument of the call is assigned to the $i$-th
formal parameter of the method.

In terms of declarative computation, this rule is evaluated via a
relational join of the current contents of (conceptual) tables
\sv{CallGraphEdge}, \sv{FormalParam}, and \sv{ActualParam}. But it is
up to the compiler to decide whether to start the join from table
\sv{CallGraphEdge} or one of the others. This decision may be informed
by dynamic statistics---i.e., by current knowledge of the size of each
of the three tables and of the past selectivity of joining each two
tables together. It could well be that our input consists of
overwhelmingly zero-argument functions. Thus, the join of
\sv{CallGraphEdge} and \sv{FormalParam} will be small. It is wasteful
(up to asymptotically so) to start by iterating over all the contents
of \sv{CallGraphEdge}, only to discover that most of them never
successfully match a method with an entry in table
\sv{FormalParam}. Instead, the join may be much quicker if one starts
from functions that do take arguments, i.e., from table
\sv{FormalParam}. The LogicBlox Datalog engine~\cite{Aref:2015:DIL:2723372.2742796}
performs precisely this kind of dynamic, online query optimization,
based on relation sizes and expected selectivities.

\paragraph{Example: Choice of Data Structures.}
Data structure choice is already standard practice in relational
languages. For instance, the \souffle{}~\cite{Jordan16} implementation
of Datalog automatically infers when to add indexes to existing
tables, so that all rule executions are
fast~\cite{DBLP:journals/pvldb/SuboticJCFS18}. In our earlier example, \souffle{} will
add an index (i.e., a B-tree or trie) over table \sv{FormalParam},
with the second column, \sv{method}, as key, and similarly for
\sv{ActualParam}, with either column as key. Then, if computation
starts from an exhaustive traversal of \sv{CallGraphEdge}, only the
matching subsets of the other two tables will be accessed, using the
index to identify them. We illustrate below, by denoting the partial,
indexed traversal by a $\Pi$ prefix on the accessed-by-index tables,
and by underlining the variables bound by earlier clauses during the
evaluation:

\noindent\begin{minipage}{\columnwidth}
\begin{datalogcode}
Assign(formal, actual) :-
  CallGraphEdge(invocation, method),
  (*$\Pi$*)FormalParam(index, (*\underline{\texttt{method}}*), formal),
  (*$\Pi$*)ActualParam((*\underline{\texttt{index}}*), (*\underline{\texttt{invocation}}*), actual).
\end{datalogcode}
\end{minipage}

Note that such choice of data structure is not based on local
constraints, but on all uses of the table, in any rule in a
(potentially large) program. However, per our discussion of trends, it
is typically fine for the compiler to maintain an extra data
structure, if this will turn an exhaustive traversal into an indexed
traversal, even if the benefit arises in very few rules.

Generally, I believe it is almost a foregone conclusion that
next-paradigm programming languages will perform automatic data
structure selection. The language will likely only require the
programmer to declare data and will then automatically infer efficient
ways to access such data, based on the structure of the
computation. Both technology trends and data structure evolution
conspire to make this scenario a near certainty:

\begin{itemize}[leftmargin=12pt,itemsep=2pt]
\item Although many potential data structures exist, a
  logarithmic-complexity, good-locality, ordered structure (such as a
  B-tree or trie) offers an excellent approximation of most realistic
  data traversals. Both random access and ordered access are
  asymptotically fast, and constant factors are
  excellent. (Accordingly, most scripting languages with wide adoption
  in recent decades have made a standard ``map'' their primary
  data type.)
  If one adds a union-find tree, abstracted behind an ``equivalence
  class'' data type, there may be nearly nothing more that a
  high-productivity language will need for the vast majority of
  practical tasks.

  Of course, the are glaring exceptions to such broad
  generalizations---e.g., there is no provision for probabilistic data
  structures, such as bloom filters, cryptographically secure
  structures, such as Merkle trees, or other classes of
  structures essential for specific domains. However, the use of such
  structures is substantially less frequent. Additionally, a theme for
  next-paradigm languages will be escaping the language easily---as I
  argue later (Section~\ref{sec:others}).
  
\item Adding an extra data structure vs. not adding a data structure
  is no longer a meaningful dilemma, under current memory and speed
  trends. The cost of additional ways to organize data only grows
  linearly, while the speed benefit can be asymptotic. Therefore, when
  in doubt, adding an extra B-tree or trie over a set of data is an
  easy decision.
\end{itemize}



\paragraph{Example: Auto-Incrementalization.} Another realistic example
of asymptotic complexity improvements offered routinely in declarative
languages is automatic incrementalization. Our earlier example rule
is, in practice, never evaluated as a full join of tables
\sv{CallGraphEdge}, \sv{FormalParam}, and \sv{ActualParam}. The reason
is that other rules in a typical program analysis will use the
resulting relation, \sv{Assign}, in order to infer new call-graph
edges (e.g., in the case of virtual calls). This makes the computation
of \sv{Assign} mutually recursive with that of \sv{CallGraphEdge}.
Therefore, the rule will be evaluated incrementally, for each stage
of recursive results. The rule, from the viewpoint of the Datalog compiler
looks like this:

\begin{datalogcode}
(*$\Delta$*)Assign(formal, actual) :-
  (*$\Delta$*)CallGraphEdge(invocation, method),
  FormalParam(index, method, formal),
  ActualParam(index, invocation, actual).
\end{datalogcode}

This means that the new-stage (denoted by the $\Delta$ prefix)
results of \sv{Assign} are computed by joining only the newly-derived
results for \sv{CallGraphEdge}. Tuples in \sv{CallGraphEdge} that
existed in the previous recursive stage do not need to be considered,
as they will never produce results not already seen. (The other two
tables involved have their contents fixed before this recursive
fixpoint.)  In practice, such automatic incrementalization has been a
major factor in making declarative implementations highly
efficient---often much faster than hand-written solutions, since
incrementalization in the case of complex recursion is highly
non-trivial to perform by hand.

Incrementalization also exhibits complex interplay with other
algorithmic optimizations. For instance, the latest delta of a
table is likely smaller than other relations, in which case the exhaustive
traversal of a join should start from it.

\begin{corollary}[Cheapest is hardest.]
\label{cor:cheap}
  ``Easy'' in terms of (sequential) computational complexity may mean
  ``hard'' to express efficiently in next-paradigm languages.
\end{corollary}

The shortcomings of next-generation languages may be more evident in
the space where human ingenuity has produced incredibly efficient
solutions, especially in the low-end of the computational complexity
spectrum (i.e., linear or near-linear algorithms). In the
ultra-efficient algorithm space, there is much less room for automated
optimization than in more costly regions of the complexity
hierarchy.\footnote{This general conjecture may be easily violated in
  specialized domains where symbolic search already \emph{beats} human
  ingenuity. E.g., program synthesis has already exhibited remarkable
  success in producing optimal algorithms based on bitwise
  operators~\cite{Gulwani:2011:SLP:1993498.1993506}.}

\paragraph{Example: Depth-First Algorithms and Union-Find Structures.}
Current declarative languages are markedly bad at expressing (without
asymptotic performance loss) efficient algorithms based on depth-first
traversal.  For instance, declarative computation of
strongly-connected components in directed graphs is asymptotically
less efficient than Tarjan's algorithm.  Also, union-find trees cannot be
replicated and need special-purpose coding.

Generally, algorithms that are hard to parallelize (e.g., depth-first
numbering is $P$-hard) and data structures that heavily employ
imperative features (both updates and aliasing) are overwhelmingly the
ones that are a bad fit for declarative programming. It is reasonable
to speculate that this observation will generalize to any
next-paradigm programming language. After all, a high-productivity
language will need to be abstract, whereas imperative structures and
non-parallelizable algorithms rely on concrete step ordering and
concrete memory relationships (i.e., aliasing). If this speculation
holds, it is a further argument for the inevitability of next-paradigm
programming languages. In most foreseeable technological futures,
parallelism and non-random-access memory are much more dominant than
sequential computation and a shared, random-access memory space.  The
algorithms that will dominate the future are likely amenable to
general automatic optimization in a high-productivity language.

\begin{corollary}[Even Asymptotics May Not Matter]
  Asymptotically sub-optimal computations may become dominant,
  for limited, well-supervised deployment.
\end{corollary}

Asymptotic performance degradation factors are impossible to ignore,
since they typically turn a fast computation into an ultra-slow or
infeasible one. However, in next-paradigm languages, a programmer may
routinely ignore even asymptotic factors and favor ultra-convenient
programming. To avoid performance degradation in a realistic setting,
the applicability of inefficient computations may be limited to a
local setting, or approximate results may be acceptable~\cite{DBLP:conf/pldi/CarbinKMR12}.

\paragraph{Example: Inefficient Graph Computations.} In Datalog code
I have often favored quadratic, cubic, or worse solutions, as long as
they are applied only locally or other constraints ensure efficient
execution. Graph concepts offer generic examples. (In practice the
computation is rarely about a literal graph, but binary relations are
often convenient viewed in graph terms.) For instance, I have often
used code that computes all pairs of predecessors of a graph node,
generically written as:

\begin{datalogcode}
BothPredecessors(pred1, pred2, next) :-
  Edge(pred1, next),
  Edge(pred2, next),
  pred1 != pred2.
\end{datalogcode}

As long as the in-degree of the graph is bounded, the ``wasteful''
all-pairs concept costs little to compute and can be quite handy
to have cached.

Similarly, a wasteful but convenient concept is that of
directed graph reachability without going through a given node:

\begin{datalogcode}
ReachableExcluding(node, node, notInPath) :-
  IsNode(node),
  IsNode(notInPath),
  node != notInPath.
  
ReachableExcluding(source, target, notInPath) :-
  Edge(source, target),
  IsNode(notInPath),
  source != notInPath,
  target != notInPath.

ReachableExcluding(source, target, notInPath) :-
  ReachableExcluding(source, interm, notInPath),
  Edge(interm, target),
  target != notInPath.
\end{datalogcode}

Note that the computation is worst-case bounded only by a $n^4$
polynomial, for $n$ graph nodes---e.g., the last rule enumerates
near-all possible node 4-tuples, \sv{source}, \sv{target},
\sv{interm}, and \sv{notInPath}.

Written as above, the computation would be infeasible for any but the
smallest graphs. However, if we limit our attention to a local
neighborhood (for whatever convenient definition, since this pattern
applies in several settings) the computation is perfectly feasible,
and, in fact, common in production code:

\begin{datalogcode}
ReachableExcluding(node, node, notInPath) :-
  InSameNeighborhood(node, notInPath),
  node != notInPath.

ReachableExcluding(source, target, notInPath) :-
  Edge(source, target),
  InSameNeighborhood(source, target),
  InSameNeighborhood(source, notInPath),
  source != notInPath,
  target != notInPath.

ReachableExcluding(source, target, notInPath) :-
  ReachableExcluding(source, interm, notInPath),
  Edge(interm, target),
  InSameNeighborhood(source, target),
  target != notInPath.
\end{datalogcode}

Generally, I believe that programmers will be quite inventive in
reshaping a problem in order to employ ultra-high-level but
inefficient computations. Coding simplicity and correctness clarity
are excellent motivators for questioning whether a full, exact answer
is strictly needed.

%

\begin{corollary}[Implicit Parallelism]
  In any high-productivity setting, parallelism will be pervasive but implicit.
\end{corollary}

A next-paradigm programming language, offering orders-of-magnitude
productivity improvements, will very likely heavily leverage
parallelism, yet completely hide it! There is no doubt that
shared-memory concurrency correctness is among the thorniest
programming challenges in existence. High-productivity and explicit
synchronization, of any form, are very unlikely to be compatible.
High levels of abstraction also seem to mesh well with automatic data
partitioning and replication solutions, as does the earlier observation
about sacrificing even asymptotic optimality on the altar of programming
productivity.

\paragraph{Example: Implicit Shared-Memory Parallelism.} Declarative languages
already auto-parallelize programs. All the Datalog logic we have seen
is executed by a modern engine (e.g., \souffle{}) in parallel, by
multiple threads over partitionings of the input data. A join
operation is naturally massively parallel, so, the larger the input
data, the more parallelism one can easily attain. 


\subsection{Programming Model: Semantic Invariants}

In addition to a cost model, a next-paradigm PL programmer's mental
model should include semantic guarantees.

\begin{principle}[Need For Firm Mental Grounding]
The programming model of next-paradigm languages will offer strong
semantic guarantees (about what code can do and how new code can affect old).
\label{principle:grounding}
\end{principle}

A language that will offer orders-of-magnitude improvements in
productivity will necessarily enable the programmer to express more
with less. A highly concise specification will yield a detailed
optimized implementation, with the compiler playing a huge role in
searching the space of potential algorithms and data
structures. Keeping one's sanity will not be trivial. A one-word (or
even one-character) change has the potential to completely alter
program output or its performance. I am not giving examples from my
declarative programming experience because anecdotes don't do justice
to the magnitude of the issue. After all, small changes with large
effects can also arise in conventional programming practice. However,
what is currently an extreme case will be common, everyday experience
in next-paradigm languages. Changes that the programmer considers
completely innocuous may result in vastly different program
implementations.

Faced with such complexity, the language will need to provide firm,
sanity-keeping semantic guarantees. What can these guarantees be?  It
is hard to speculate on specifics without knowing the exact technology
behind a language. Most likely, semantic invariants will guarantee
what the program can or cannot do, what effect code in one part of the
program can have on others, how the program output can change once new
code is added, etc. Current examples give a glimpse of the
possibilities.

\paragraph{Example: Monotonicity.}
The top sanity-preserving semantic invariant in Datalog is
monotonicity. Computation is (dominantly) monotonic---other rules can
only \emph{add} to the output of a rule, never invalidate previous
outputs. This gives excellent properties of code understanding via
local inspection. It also helps with understanding the updated program
semantics upon a code change---code addition is overwhelmingly the
most common change in program development. Consider the earlier
example of an \sv{Assign} inference from established instances of
\sv{CallGraphEdge}. To understand the rule, we have never needed to
wonder about either other rules that inform \sv{Assign} or the rules
that establish \sv{CallGraphEdge}, even if those themselves employ
\sv{Assign}. Also, we have never needed to wonder about the evaluation
order of this rule relative to any others. The rule works like a pure
logical inference precisely because of the monotonicity property. The
same holds for the three rules we employed for
\sv{ReachableExcluding}: we have not needed the definition of any rule
in order to understand the others.\footnote{In reality, Datalog
  programs typically use ``stratified negation'', allowing
  non-monotonic operators (i.e., negation and arbitrary aggregation)
  over predicates defined in a previous evaluation stratum.  This
  means that extra rules \emph{can} produce fewer results, but only if
  these extra rules affect a predicate that is conceptually ``more
  primitive'' than the extra results. The programmer needs to have a
  stratification of the program logic in mind. This is a fairly
  manageable mental burden (given compiler support for tracking
  violations), even for large code bases.}




\paragraph{Example: Termination.}
A guarantee of program termination is another semantic invariant of
the Datalog language. Core Datalog does not include operators for
inventing new values, therefore all computation can at most combine
input values. This results in a polynomial number of possible
combinations, more and more of which are computed monotonically during
program evaluation. Therefore, a program is guaranteed to terminate.

Of course, in practice strict guarantees of termination are impossible
or impractical. A guarantee of termination makes a language sub-Turing
complete. Practical implementations of Datalog allow creating new
values, thus no ironclad guarantee of termination exists. However, it
is valuable to distinguish a core guarantee, applicable to the
majority of the code, from an exception that can be effected only by
use of a well-identified language feature.\footnote{``Pure''
  functional languages do something similar. They certainly permit
  side effects (e.g., I/O), but encapsulate them behind a
  well-identified interface---a monad.}  It is much easier to reason
about the possible new values invented via constructors than have
potentially non-terminating computations possibly lurking behind every
computation in the language.

\subsection{``Known'' Principles? \\
  Abstraction, Extensibility, Modularity, Interoperability}
\label{sec:others}

It is interesting to speculate on new design principles of
next-paradigm languages, but what about current, well-established
language principles?  These include, at the very least:

\begin{itemize}[leftmargin=12pt,itemsep=2pt]
\item (module) abstraction/growing a language: packaging recurring
  functionality in a reusable, parameterizable
  module~\cite{Steele:1998:GL:346852.346922};
\item language extensibility: the marriage of powerful module
  abstraction with syntactic configurability;
\item modularity: having mechanisms for keeping parts
of the code isolated, only visible through specific, identified
interfaces;
\item multi-paradigm interoperability: easy interfacing between
  languages so that programming tasks can be expressed in terms
  well-suited to the computation at hand.
\end{itemize}

There is no doubt that current, established principles of good
language design will play a big role in next-paradigm language design
as well. However, these principles on their own are not enough to get
us to a next-paradigm language. Furthermore, the benefits the
established principles afford may be only second-order effects. They
may pale compared to the chief benefit of a next-paradigm language:
levels of abstraction high enough to yield orders-of-magnitude
productivity improvements. In fact, the first next-paradigm
programming languages may well be lacking in module abstraction,
extensibility, or modularity. Later incarnations will likely benefit
more from these principles, as have existing languages.

The one principle that seems a priori indispensable for next-paradigm
languages is that of multi-paradigm interoperability. I already
conjectured that next-paradigm programming languages will not be good
at everything---e.g., see Corollary~\ref{cor:cheap}. Escaping the
language is then inevitable, even if it happens rarely. The language
escape mechanism should not break the fundamental high-productivity
abstraction. Instead, it can encapsulate, behind a clear interface,
computations best optimized by manual, lower-level coding. There are
mechanisms of this nature in common current use---e.g., uninterpreted
functions/external functors.

\paragraph{Diminishing Returns.}
The need to escape a language raises an interesting question. How can
high productivity be sustained, if parts of the coding need to happen
in a conventional setting? The law of diminishing returns dictates
that dramatically speeding up one part of software development will
only move the bottleneck to a different part. If, say, the core of
application logic currently takes $70\%$ of programming effort and is
made $10$x more efficient, with other tasks staying the same, then
$37\%$ of the original effort is still required---the overall benefit
is merely a factor of $2.7$x.  Tasks that may not benefit much from
next-paradigm languages include low-level coding, as well as bespoke
code for UI/storage/other interfacing with the environment.

Indeed, this is a constraint that can limit the large benefits of
high-productivity languages to some kinds of development tasks but not
others. However, I believe the effect will be largely mitigated by
more conventionally forms of productivity enhancement: domain-specific
languages and reusable modules (i.e., module
abstraction). Furthermore, many of the non-core tasks of program
development parallelize a lot more easily than the core application
logic itself. Building a UI, a service wrapper, or integration APIs
may require effort, but the effort can be more easily split among
multiple programmers. Therefore, even though the overall effort may
indeed see smaller (than orders-of-magnitude) improvements, other
metrics of development productivity, such as end-to-end development
time, may improve more, by dedicating more programmers to the task.


\vspace{0.5cm}

\epigraph{We have programs that are vastly powerful but also vastly mysterious, meaning small changes can badly destabilize the system and we don't yet know how to talk about debugging or maintenance.}{\textit{Jan-Willem Maessen, in reference to Sculley et al.~\cite{43146}}}

\section{Changes to Development Workflows}

An informal axiom of the Software Engineering community
is that the choice of programming language does not fundamentally
change the software engineering \emph{process}. This may be true in
the sense that process stages (e.g., requirements analysis,
architecture, design, coding, testing, verification) remain the same
conceptual entities. However, the relative effort and emphasis of each
stage may change dramatically with high-productivity, high-abstraction
languages.  The \emph{practice} of software development will change.

\begin{principle}[Workflows Will Change]
  Next-paradigm programming languages will change well-established
  patterns in current programming workflow.
\label{principle:workflow}
\end{principle}

Orders-of-magnitude productivity improvements will, very likely,
disrupt the established workflow of program development. Code will be
much more terse and abstract, resembling a formal specification.
Small changes will have a huge impact on both functionality and
performance. It is hard to fully grasp precisely how today's common
practices will evolve, so I will speculate modestly, straying little
from observations from my own coding practices.

\begin{corollary}[Incremental Development]
  In next-paradigm languages, it will be much easier to develop programs
  incrementally, and to continue from where one has left off.
\end{corollary}

This observation may seem pedestrian, but it has been one of the most
striking revelations in my everyday experience with declarative
programming. The higher level of abstraction means that one can rely
on highly-powerful concepts without ever looking at their
definitions. Specifically for Datalog development, monotonicity in the
language evaluation model means that developing more functionality is
a natural extension of what is already there. However, the experience
of incremental development will likely generalize to all higher-level
programming settings. Programming with high-level specifications is
naturally incremental. For one, at any point in development, a partial
specification yields a program. The program's outputs are incomplete,
but they immediately suggest what is missing and how more work can get
the program closer to the desired task. Adding extra features
interacts in a predictable way with earlier functionality.

\paragraph{Example: An Interactive Graphical Application.} Imagine
creating an interactive graphical application (e.g., a video game or a
drawing program) in a language with high degrees of abstraction. Such
a language will likely accept a logical specification of \emph{what}
should be displayed, \emph{where} and \emph{when}, without encoding at
all the \emph{how}. Whether a control should have an effect at a
certain point, whether a graphical element is visible or obstructed,
how the display adjusts to changes of the environment, etc., are all
elements that the language implementation should address, and not the
program itself.

Development in such a language is naturally incremental. It is
straightforward to start with specifications of elements that appear
on the screen, under some control conditions (which may involve
timing). There is no need to specify yet what the timing or the
controls are---just to declare them. It is easy to add sample inputs
for these and see graphical elements displayed---the incomplete
specification already yields a working program. Making progress on any
front is incremental and straightforward. One can add more graphical
elements, or more detail over current elements, or complex
specifications to define user control, or a mechanism for storing
data, or any other desired functionality. All new functionality should
interact cleanly with existing functionality. The language
implementation will resolve conflicts (sometimes with the programmer's
help, in case of hard conflicts) and will produce vastly different
low-level programs as the specification becomes more and more
complete.


\begin{corollary}[Testing Changes]
  In next-paradigm languages, testing will be a larger and deeper part
  of the programmer's workflow.
\end{corollary}

When programming effectively becomes a collaboration between a
creative human and a compiler with vast abilities in exploring an
implementation space, the role of testing will change dramatically.
The programmer will write highly concise code that produces very
complex outputs. Continuous checking of assumptions against the
implemented model will be necessary. The programmer may be spending
much more time running code in complex settings than writing it.

The task of testing and debugging will also be conceptually
harder. Testing may be as complicated as writing the code, and indeed
writing testing specifications may be obligatory. The difficulty
arises because efficiency in execution necessarily means that most
intermediate results will never be part of the output. This
complicates debugging enormously. For instance, the mechanism of
time-travel debugging (e.g., see
Reference~\cite{Barr:2014:TAT:2660193.2660209}), which has captured
the programming community's imagination in recent years, works only
when the space of intermediate values of a computation is not that
much larger from the space of final values. This is not the case for
abstract programs. Both the language implementation and the program
itself may be collapsing a much larger space of values in order to get
a single output~\cite{DBLP:conf/datalog/KohlerLS12}.

\paragraph{Example: Paths.} A pedestrian but illustrative example is
that of a declarative ``transitive closure'' computation: compute when
there is a path between two nodes, given direct edges as input. The
recursive rule for this is:

\begin{datalogcode}
Path(source, target) :-
  Edge(source, interm),
  Path(interm, target).
\end{datalogcode}

This computation only stores the fact that there \emph{is} a path, but
not how this path was established: the \sv{interm} value is dropped.
Keeping all \sv{interm} values will make the cost of computing and
storing the \sv{Path} relation combinatorially larger: $O(n^3)$
instead of $O(n^2)$.\footnote{If one wants to be pedantic, in the
  worst case, the cost of computing \texttt{Path} is $O(n^3)$
  anyway. But in the sparse graphs that arise in practice, the
  computation is typically $O(n^2)$ if intermediate nodes do not need
  to be kept and $O(n^3)$ if they do.}  In practice, this increase is
often prohibitive. Consider that the transitive closure computation
shown in the above rule is the simplest possible recursive computation
in a logical specification. Most real specifications that employ
recursion will be much more complex, with several intermediate values
used to derive results but not memorized. Therefore, completely
routine computations become intractable if it becomes necessary to
fully trace how the result was derived.

\begin{corollary}[Different Balance of Formal Reasoning and Coding]
  For programs in next-paradigm languages, formal proofs will be
  easier.  Yet they will also be less necessary.
\end{corollary}
A program that is effectively an executable specification removes some
of the need for formal verification. The value of verification in
current coding stems partly from the assurance of formal reasoning and
partly from specifying the computation in a completely different
formulation, unencumbered by implementation constraints. In a
next-paradigm language, the level of abstraction of the program will
likely be much closer to that of a non-(efficiently-)executable
specification. There will likely still be a need for formal reasoning,
to establish properties of the program with confidence and in full
generality. Such reasoning will be easier just by virtue of the
smaller gap between the two specifications.

\section{Conclusions}

Next-paradigm programming languages will need a revolutionary change
in level of abstraction, if they are to ever realize large
productivity improvements. Such a change will necessarily have many
repercussions on the role of the compiler, on the programmer's mental
model, and on development patterns. In this paper, I tried to identify
these changes, looking over the future through the misty glass of the
present. Necessarily, I only present the view of the future from where
I currently stand. Others will likely find it too speculative or too
myopic---and that's fine. But we need a conversation about
next-paradigm programming languages and I hope to help start it.

\bibliography{bibliography,references,bib/ptranalysis,bib/proceedings,bib/specs,bib/tools}



\end{document}